\documentclass{article}
\usepackage{spconf,amsmath,graphicx,tabularx}

\usepackage{enumitem}
\usepackage{inputenc}
\setlist{nosep, leftmargin=14pt}

\usepackage{mwe} 

\usepackage{bibspacing}
\def\L{{\cal L}}

\title{Automatic Detection of B-lines in Lung Ultrasound Videos From Severe Dengue Patients}
\name{%
\begin{tabular}{@{}c@{}}
Hamideh Kerdegari$^{a}$ \quad Phung Tran Huy Nhat$^{a, b}$ \quad Angela McBride$^{b}$\quad VITAL Consortium$^{e}$\\
Reza Razavi$^{a}$\quad Nguyen Van Hao$^{c}$\quad Louise Thwaites$^{b, d}$\quad Sophie Yacoub$^{b, d}$ \\
Alberto Gomez$^{a}$
\end{tabular}\thanks{This work was supported by the Wellcome Trust UK (110179/Z/15/Z, 203905/Z/16/Z). H. Kerdegari, N. Phung, R. Razavi and A. Gomez also acknowledge financial support from the Department of Health via the National Institute for Health Research (NIHR) comprehensive Biomedical Research Centre award to Guy's and St Thomas' NHS Foundation Trust in partnership with King's College London and King's College Hospital NHS Foundation Trust.}}

\address{$^{a}$ School of Biomedical Engineering \& Imaging Sciences, King's College London, UK \\
     $^{b}$Oxford University Clinical Research Unit, Ho Chi Minh City, Vietnam \\
     $^{c}$Hospital for Tropical Diseases, Ho Chi Minh City, Vietnam \\
     $^{d}$Centre for Tropical Medicine and Global Health, University of Oxford, UK \\
     $^{e}$a corporate author; for the list of members, please see the section at the end of
this manuscript}
\begin{document}
%
\maketitle
\begin{abstract}
Lung ultrasound (LUS) imaging is used to assess lung abnormalities,
including the presence of B-line artefacts due to
fluid leakage into the lungs caused by a variety of diseases.
However, manual detection of these artefacts is challenging.
In this paper, we propose a novel methodology to automatically
detect and localize B-lines in LUS videos using deep
neural networks trained with weak labels. To this end, we
combine a convolutional neural network (CNN) with a long
short-term memory (LSTM) network and a temporal attention
mechanism. Four different models are compared using data
from 60 patients. Results show that our best model can determine
whether one-second clips contain B-lines or not with an
F1 score of 0.81, and extracts a representative frame with B-lines with an accuracy of 87.5\%.
\end{abstract}
\section{Introduction}
\label{sec:intro}

Ultrasound imaging is gaining popularity for real-time patient management in the intensive care units (ICU) because it is mobile, fast, non-invasive, safe for patients and relatively inexpensive. Specifically, lung ultrasound (LUS) is becoming the reference modality for rapid lung assessment but unlike all other ultrasound imaging applications, the purpose of LUS is to capture image artefacts that indicate a pulmonary abnormality, including features of extravascular lung water such as oedema and effusions \cite{soldati2019ultrasound}. Fluid leakage is one of the characteristic clinical features of severe dengue and accurate assessment of this is critical for dengue patient care \cite{mayo2019thoracic}. To this end, ultrasound imaging can be used to assess leakage through the presence and appearance of B-lines (e.g. Fig.1, frame 12), bright lines extending from the surface of the lung distally following the direction of propagation of the sound waves. These lines appear and disappear during the respiratory cycle and may be found only in some regions of the affected lung \cite{dietrich2016lung}. As a result, manually detecting these lines is a very challenging task, particularly for inexperienced operators.

\begin{figure*}[bht!]
\includegraphics[width=\linewidth]{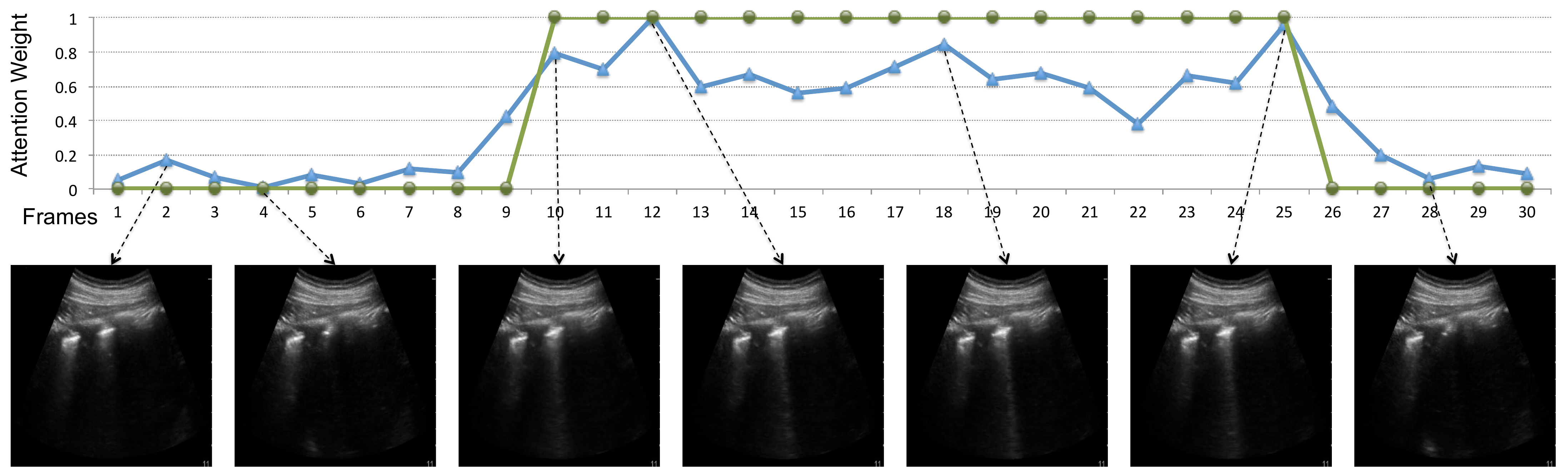}
\caption{Results of attention weights (blue line) and their associated ground truth (green line) on a sample one second LUS video containing non-B-line and B-line frames. Some example frames with their related attention weights are visualized. For example, frame 12 is a B-line frame and has an attention weight of \textit{1} while frame 2 that shows a non-B-line frame has received an attention weight of \textit{0.2}. Here, the attention weights are normalized between \textit{0} and \textit{1} using min-max normalization for visualization.}
\label{fig:att-result}
\end{figure*}

Recent advances in computer vision, machine learning and particularly deep learning have brought great advances
in challenging computer vision tasks such as classification and object detection. Applied to medical imaging, these tasks could help automate problems such as B-line detection in LUS. Despite the wide use of such techniques to more common applications of ultrasound imaging, very few works, and very recently only, have been published on automatic analysis of LUS images. Related work can be organised in two categories: detection of B-lines (or other artefacts) in an image; and segmentation or localisation of lung lesions in an image.

The first category of methods use classification techniques to detect B-lines in individual LUS frames. For example,
Sloun et al. \cite{van2019localizing} applied classification and weakly-supervised localization of B-lines to LUS from COVID-19 patients. A fully convolutional network was trained to recognize abnormality in images, followed by class activation maps (CAMs) \cite{zhou2016learning} to produce a weakly-supervised segmentation map of the input. Unlike \cite{van2019localizing} that used CAMs for localization, Roy et al. \cite{roy2020deep} exploited a spatial transformer network for weakly supervised localization. Further, they proposed an ordinal regression to predict the presence of COVID-19 related artefacts and a score connected to the disease severity. In another study \cite{kulhare2018ultrasound}, a single-shot CNN was employed to predict bounding boxes for B-lines. All these methods classify one frame at a time, either requiring a method to extract a frame from the ultrasound stream first, or needing a method to unify a prediction from all predictions done on individual frames from one patient.

The second category of methods is focused on using attention mechanisms, particularly on CT and x-ray lung images. A residual attention U-Net for multi-class segmentation of COVID-19 Chest CT images was proposed by Chen et al. \cite{chen2020residual}. Similar architecture was applied by Gaal et al. \cite{gaal2020attention} but for x-ray lung segmentation of pneumonia. In \cite{ouyang2020dual}, a 3D CNN network with online attention refinement and dualsampling strategy was developed to distinguish COVID-19 from the pneumonia in chest CT images. A lesion-attention deep neural network (LA-DNN) was proposed by \cite{liu2020online}, learning two tasks: a primary binary classification task on presence of COVID-19 and an auxiliary multi-label attention learning task on five lesions. It was shown that the auxiliary task promotes the primary task to focus attention on the lesion areas and consequently improve the classification performance.

In all the mentioned studies, CAMs and attention mechanisms have been used for the spatial localization of lung lesions. Differently, we leverage temporal analysis networks and use attention to find and localize the most important frames (i.e. B-line frames) within a video that contains B-lines. Indeed, B-line artefacts appear at arbitrary frames within a LUS video, hence the ability to first detect whether there are B-line frames in the video or not is essential for clinical applicability. A variety of classical models \cite{sminchisescu2006conditional, ikizler2007searching} have been applied for temporal context modeling. Most recently, RNNs and particularly LSTM have become popular due to their ability for end-to-end training when combined with CNN. Several recent studies incorporated spatial/optical-flow CNN features with LSTM models for global temporal modeling of videos \cite{srivastava2015unsupervised, donahue2015long}. We also incorporate CNN features with LSTM for LUS video classification. However, we use a new variant of the LSTM model equipped with an attention network that allows it to focus and highlight B-lines artefacts as discriminative frames in the ultrasound video.

In summary, the novel contributions of this paper are: 1) analysis of ultrasound videos, instead of ultrasound frames, exploiting temporal information that captures the dynamic nature of the underlying anatomy; and 2) utilization of temporal attention to localise in time the video frames where B-lines are shown.

\section{Model Architecture}
\label{sec:Method}

\begin{figure*}[htb]
\includegraphics[width=\linewidth, trim={0 7 0 3}, clip=true]{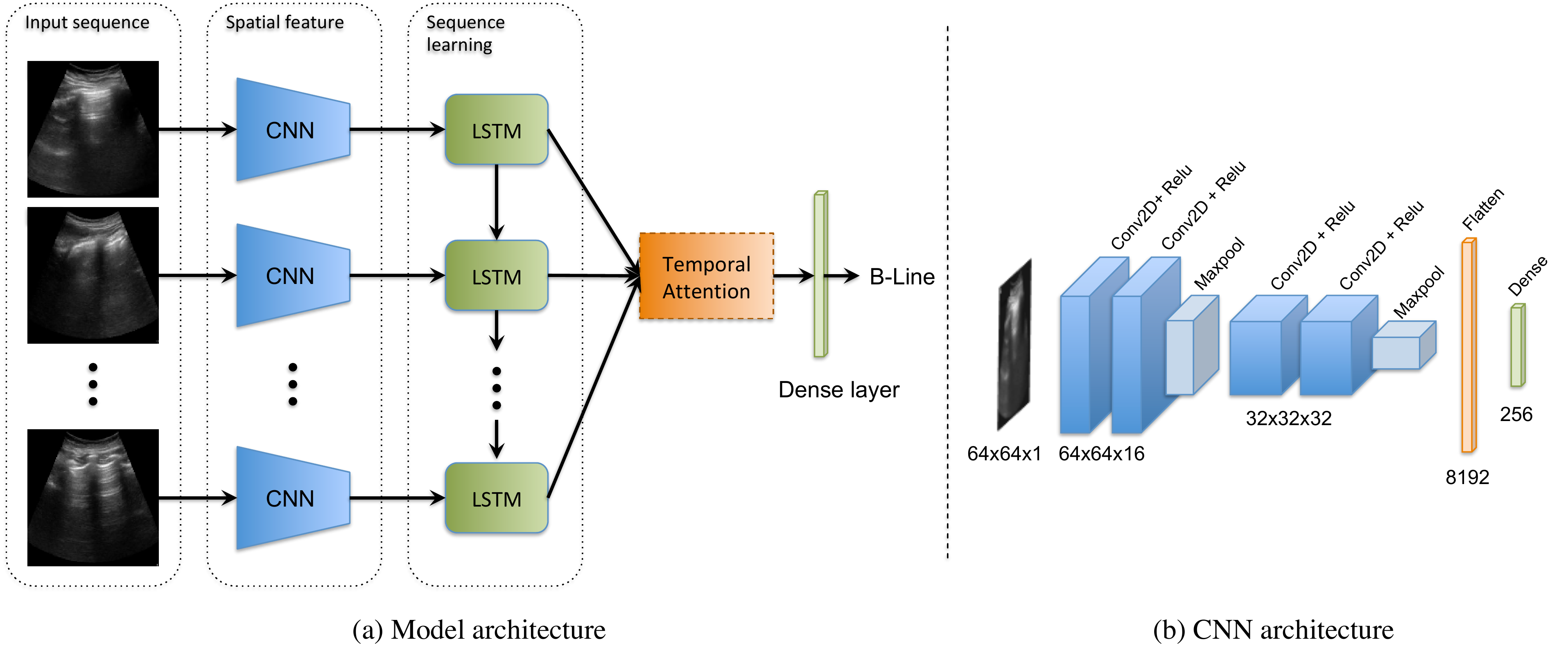}
\caption{The proposed architecture for LUS B-line detection. (a): This model consists of CNN layers, a bidirectional LSTM, and an attention module. (b): The detailed architecture of the CNN model.}
\label{fig:model-arch}
\end{figure*}

The overall model architecture is shown in Fig.\ref{fig:model-arch}a. It consists mainly of three parts: convolutional neural network (CNN), bidirectional long short-term memory (LSTM) network and temporal attention mechanism. 

The input to the model is a sequence of N frames that is represented using a matrix of $X = (x_{0}, ...,x_{N}), X \in R^{D}$. Spatial features are extracted from this sequence using the CNN model described below.  The CNN architecture (shown in Fig.\ref{fig:model-arch}b) consists of four layers of convolution with \emph{ReLU} activation, and two max-poolings. Each convolution filter uses $3 \times 3$ kernels with unit stride. A fully connected layer is used at the end to produce a 256-dimensional feature vector to represent each frame in the input video. Then, this feature vector is passed as input to the bidirectional LSTM to extract temporal features. We use a bidirectional LSTM with 16 hidden units and \emph{tanh} activation function. The LSTM outputs are then passed to the attention network to generate an attention score. We adopt the temporal attention mechanism proposed by Bahdanau et al. \cite{bahdanau2014neural} for neural machine translation. Specifically, this attention model computes an attention score $e_{t}$ for each attended frame $h$ at time step $t$: 

\begin{equation}
e_t=h_t w_a
\end{equation}
Here $h_{t}$ is the representation of the frame at time step t and $w_{a}$ is the weight matrix for the attention layer. From the attention score $e_t$, an importance attention weight  $a_{t}$ is computed for each frame at each time $t$:
\begin{equation}
a_t = \frac{\exp (e_{t})}{\sum_{i=1}^{T}\exp (e_{i})}
\end{equation}

The importance attention weights are multiplied by the feature vector output by the LSTM, hence effectively learning which frame of the video to pay attention to. A higher attention weight reflects a more discriminative value of the frame with respect to the B-line detection task. The attention-weighted temporal feature vector is averaged over time, $\bar{A} =\frac{1}{n}\sum_{i=1}^{n}A_{i}$ and passed to a fully connected layer for the final LUS video classification.

\section{Data, Materials and Experiments}
\label{sec:Experiments}
In this section, the data collection procedure and materials used are first explained. Then, experiments and evaluation criteria are presented. 

\subsection{Data}
\label{ssec:Dataset Acquisition}

The LUS exams were carried out using a Sonosite M-Turbo machine (Fujifilm Sonosite, Inc., Bothell, WA) with a low-medium frequency (3.5-5 MHz) convex probe by qualified sonographers. LUS was performed using a standardised operating procedure based on the Kigali ARDS protocol \cite{riviello2016hospital}: assessment for B-lines \cite{lichtenstein2008relevance,volpicelli2012international}, consolidation and pleural effusion, performed at 6 points on each side of the chest (2 anterior, 2 lateral and 2 posterolateral).  

For this study, data from 60 patients were acquired between June 2019 and June 2020. Each patient had an average five LUS examinations, totaling 298 examinations. The video resolution was $\ 640 \times 480$ with a frame rate of 30fps. The acquired dataset has about five hours LUS video data containing B-line and non-B-line videos. Four-seconds clips at each acoustic window were stored as AVI\footnote{Audio Video Interleave (AVI)} format and fully anonymised through masking. These video clips were annotated by a qualified sonographer using the VGG annotator tool \cite{dutta2019vgg}. The annotation procedure was performed by assigning a label (either B-line or non-B-line) to each video clip and then localizing the B-line frames in the B-line videos. Then, the annotation output was saved in JSON\footnote{JavaScript Object Notation (JSON): https://www.json.org/} format ready to be used by the model. For the model training, each four-seconds clip was converted into shorter clips of one second with an overlap of 20 percent between consecutive frames in the video.

\subsection{Materials and Implementation Details}
\label{ssec:Experimental Setups}

The proposed model was implemented using Keras library with a Tensorflow backend. The standard Adam optimizer was used for the network optimization with the learning rate set to 0.0001. A batch size of 20 and batch normalization were utilized for both convolutional and LSTM network layers. Dropout of 0.2 and $\L2= 10^{-5}$  for regularization were considered. During the training stage, all the input videos were resized to $\ 64 \times 64$ video clips. The dataset was augmented by adding horizontally-flipped frames to the training data. We used 5-fold cross validation and trained the network for 60 epochs.

\subsection{Experiments}
As an evaluation metric for the classification task, precision, recall, and F1 score were reported. Intersection Over Union (IoU) of the predicted and ground truth temporal labels was used as the attention error metric.

To evaluate the potential benefit of exploiting temporal information and the effectiveness of the attention mechanism, four model architectures were compared: as a baseline, 2D convolutions in the initial CNN subnet followed by temporal attention module and no LSTM (C2D+A); a model with 3D convolutions in the CNN subnet followed by temporal attention and no LSTM (C3D+A); a model with 2D CNN followed by LSTM (C2D+LSTM); and last, a model with 2D CNN followed by LSTM and temporal attention (C2D+LSTM+A).

\subsection{Results}
\label{ssec:Results}

Results on our LUS video dataset are presented in Table \ref{tab:Accuracy}. As it is shown, C2D+A model has the least F1 score because it can not model the temporal aspect of the data with 2D CNN and no recursion over time. Using C3D+A model, the performance improves which shows the ability of C3D+A structure for modelling the temporal aspect of the data but with a short context span over time. However, adding LSTM to C2D model (C2D+LSTM) indicates that LSTM part of the model is crucial for the final performance as it considers long temporal progression of the LUS video data. Finally, C2D+LSTM+A model outperforms the other models and shows that with the temporal attention mechanism F1 score improved from 0.79 (in C2D+LSTM) to 0.81 (+ 0.02). This experiment demonstrates that all the sub-components of the proposed method contribute to the final performance improvement.

\begin{table}[h]
\caption{Precision, Recall and F1 score results on the LUS video dataset using different models.}
\label{tab:Accuracy}
\centering
\begin{tabular}{l|lll}
\hline
\hline
\textbf{Model}                                                             & \textbf{Precision} & \textbf{Recall} & \textbf{F1}\\ \hline
C2D+A                                                               &      0.57        &      0.61    &      0.58     \\ \hline
C3D+A                                                                  &     0.73   &    0.82  &    0.77    \\ \hline
C2D+LSTM & 0.75 & 0.85 & 0.79      \\ \hline
\textbf{C2D+LSTM+A}  &  0.76       &    0.89   &    \textbf{0.81}       \\ \hline
\hline
\end{tabular}
\end{table}

Besides improving the classification performance, it is shown that the temporal attention mechanism is able to highlight discriminative frames that contain B-lines quantitatively in Table \ref{tab:IoU}. 
\begin{table}[h]
\caption{B-line localization accuracy (\%) at different IoU $\alpha$'s.}
\label{tab:IoU}
\centering
\begin{tabular}{l|llll}
\hline
\hline
IoU       &  $\alpha$=0.1&  $\alpha$=0.2& $\alpha$=0.3& $\alpha$=0.4 \\ \hline
C2D+A &    36.2    &    31.4   &   28.6    &   22.4 \\ \hline
C3D+A &    63.3    &    61.1   &   54.0   &   45.7 \\ \hline
C2D+LSTM+A &    \textbf{67.1}    &    65.3   &   54.5    &   49.5 \\ \hline
\hline
\end{tabular}
\end{table}
These predicted temporal localized frames are compared with the ground truth annotation at different IoU thresholds, achieving an accuracy of up to 67.1\%. To illustrate the meaning of this number, the example shown in Fig.\ref{fig:att-result} had an IoU of 78\%. 
Further, the representative frame with B-lines (i.e. a frame with the highest attention weight) was identified on the test set with the accuracy of 87.5\%. This is useful to automatically provide clinicians with insight to localize B-lines in the LUS video. Temporal attention results are visualized in Fig.\ref{fig:att-result}. The figure shows a representative example of attention weight values on a sample LUS video containing B-line and non-B-line frames. Seven frame samples were picked that show a variety of attention weights. It shows that temporal attention module is able to automatically detect important frames and to avoid frames corresponding to non-B-line frames, which maybe irrelevant after we know there are B-lines in the sequence.

\section{Conclusion}
\label{sec:Conclusion}

We have proposed an attention-based convolutional$+$LSTM model capable of detecting the B-line artefacts and localizing them within LUS videos. This architecture allows us to capture features from both spatial and temporal dimensions. Further, the temporal attention mechanism enables the localization of B-line frames. The performance of this model was evaluated on our LUS video dataset and showed classification F1 score of 0.81 and B-line localization accuracy of 67.1\%. These results demonstrate the efficacy of our approach and are consistent with qualitative analysis via visual inspection of the calculated attentions, which highlight frames with the most salient B-lines in the video.
 
Future work includes investigating more accurate spatial feature extractors such as VGG19 \cite{simonyan2014very} and ResNet101 \cite{he2016deep} architectures that will likely lead to better overall performance. 
In addition, it is interesting to add a spatial attention mechanism to the model to detect B-line regions in the LUS video along with the B-line frames, which is the first step towards the quantification of the severity of the disease. Further, architectures like temporal convolutional networks \cite{lea2016temporal} that have worked well in other domains for sequence modeling could be applied to LUS video analysis.
Overall, our results on the automation of B-Line detection using LUS will assist the fluid status assessment and management of patients with dengue and other diseases, especially for users with less ultrasound expertise.

\section{Compliance with Ethical Standards}
\label{sec:ethics}

This study was performed in line with the principles of the Declaration of Helsinki. Approval was granted by the Ethics Committee of the Hospital for Tropical Diseases, Ho Chi Minh City and Oxford Tropical Research Ethics Committee.

\section{Acknowledgments}
\label{sec:acknowledgments}
A. G. is an advisor to Ultromics Ltd. The VITAL Consortium: \textbf{OUCRU}: Dang Trung Kien, Dong Huu Khanh Trinh, Joseph Donovan, Du Hong Duc, Ronald Geskus, Ho Bich Hai, Ho Quang Chanh, Ho Van Hien, Hoang Minh Tu Van, Huynh Trung Trieu, Evelyne Kestelyn, Lam Minh Yen, Le Nguyen Thanh Nhan, Le Thanh Phuong, Luu Phuoc An, Nguyen Lam Vuong, Nguyen Than Ha Quyen, Nguyen Thanh Ngoc, Nguyen Thi Le Thanh, Nguyen Thi Phuong Dung, Ninh Thi Thanh Van, Pham Thi Lieu, Phan Nguyen Quoc Khanh, Phung Khanh Lam, Phung Tran Huy Nhat, Guy Thwaites, Louise Thwaites, Tran Minh Duc, Trinh Manh Hung, Hugo Turner, Jennifer Ilo Van Nuil, Sophie Yacoub. \textbf{Hospital for Tropical Diseases, Ho Chi Minh City}: Cao Thi Tam, Duong Bich Thuy, Ha Thi Hai Duong, Ho Dang Trung Nghia, Le Buu Chau, Le Ngoc Minh Thu, Le Thi Mai Thao, Luong Thi Hue Tai, Nguyen Hoan Phu, Nguyen Quoc Viet, Nguyen Thanh Nguyen, Nguyen Thanh Phong, Nguyen Thi Kim Anh, Nguyen Van Hao, Nguyen Van Thanh Duoc, Nguyen Van Vinh Chau, Pham Kieu Nguyet Oanh, Phan Tu Qui, Phan Vinh Tho, Truong Thi Phuong Thao. \textbf{University of Oxford}: David Clifton, Mike English, Heloise Greeff, Huiqi Lu, Jacob McKnight, Chris Paton. \textbf{Imperial College London}: Pantellis Georgiou, Bernard Hernandez Perez, Kerri Hill-Cawthorne, Alison Holmes, Stefan Karolcik, Damien Ming, Nicolas Moser, Jesus Rodriguez Manzano. \textbf{King’s College London}: Alberto Gomez, Hamideh Kerdegari, Marc Modat, Reza Razavi. \textbf{ETH Zurich}: Abhilash Guru Dutt, Walter Karlen, Michaela Verling, Elias Wicki. \textbf{Melbourne University}: Linda Denehy, Thomas Rollinson.

\bibliographystyle{IEEEbib}
\bibliography{refs}

\begin{thebibliography}{10}

\bibitem{soldati2019ultrasound}
G.~Soldati et~al.,
\newblock ``Ultrasound patterns of pulmonary edema,''
\newblock {\em Annals of Translational Medicine}, vol. 7, no. 1, 2019.

\bibitem{mayo2019thoracic}
P.~Mayo et~al.,
\newblock ``Thoracic ultrasonography: a narrative review,''
\newblock {\em Intensive Care Medicine}, pp. 1--12, 2019.

\bibitem{dietrich2016lung}
C.~Dietrich et~al.,
\newblock ``Lung b-line artefacts and their use,''
\newblock {\em Journal of Thoracic Disease}, vol. 8, no. 6, pp. 1356, 2016.

\bibitem{van2019localizing}
R.~J. van Sloun et~al.,
\newblock ``Localizing b-lines in lung ultrasonography by weakly supervised
  deep learning, in-vivo results,''
\newblock {\em IEEE JBHI}, vol. 24, no. 4, pp. 957--964, 2019.

\bibitem{zhou2016learning}
B.~Zhou et~al.,
\newblock ``Learning deep features for discriminative localization,''
\newblock in {\em CVPR}, 2016, pp. 2921--2929.

\bibitem{roy2020deep}
S.~Roy et~al.,
\newblock ``Deep learning for classification and localization of covid-19
  markers in point-of-care lung ultrasound,''
\newblock {\em IEEE TMI}, 2020.

\bibitem{kulhare2018ultrasound}
S.~Kulhare et~al.,
\newblock ``Ultrasound-based detection of lung abnormalities using single shot
  detection convolutional neural networks,''
\newblock in {\em MICCAI-PoCUS}, pp. 65--73. 2018.

\bibitem{chen2020residual}
X.~Chen et~al.,
\newblock ``Residual attention u-net for automated multi-class segmentation of
  covid-19 chest ct images,''
\newblock {\em arXiv:2004.05645}, 2020.

\bibitem{gaal2020attention}
G.~Ga{\'a}l et~al.,
\newblock ``Attention u-net based adversarial architectures for chest x-ray
  lung segmentation,''
\newblock {\em arXiv:2003.10304}, 2020.

\bibitem{ouyang2020dual}
X.~Ouyang et~al.,
\newblock ``Dual-sampling attention network for diagnosis of covid-19 from
  community acquired pneumonia,''
\newblock {\em IEEE TMI}, 2020.

\bibitem{liu2020online}
B.~Liu et~al.,
\newblock ``Online covid-19 diagnosis with chest ct images: Lesion-attention
  deep neural networks,''
\newblock {\em medRxiv}, 2020.

\bibitem{sminchisescu2006conditional}
C.~Sminchisescu et~al.,
\newblock ``Conditional models for contextual human motion recognition,''
\newblock {\em CVIU}, vol. 104, no. 2-3, pp. 210--220, 2006.

\bibitem{ikizler2007searching}
N.~Ikizler et~al.,
\newblock ``Searching video for complex activities with finite state models,''
\newblock in {\em CVPR}, 2007, pp. 1--8.

\bibitem{srivastava2015unsupervised}
N.~Srivastava et~al.,
\newblock ``Unsupervised learning of video representations using lstms,''
\newblock in {\em International conference on machine learning}, 2015, pp.
  843--852.

\bibitem{donahue2015long}
J.~Donahue et~al.,
\newblock ``Long-term recurrent convolutional networks for visual recognition
  and description,''
\newblock in {\em CVPR}, 2015, pp. 2625--2634.

\bibitem{bahdanau2014neural}
D.~Bahdanau et~al.,
\newblock ``Neural machine translation by jointly learning to align and
  translate,''
\newblock {\em arXiv preprint arXiv:1409.0473}, 2014.

\bibitem{riviello2016hospital}
E.~D. Riviello et~al.,
\newblock ``Hospital incidence and outcomes of the acute respiratory distress
  syndrome using the kigali modification of the berlin definition,''
\newblock {\em Am. J. Respir. Crit. Care Med.}, vol. 193, no. 1, pp. 52--59,
  2016.

\bibitem{lichtenstein2008relevance}
D.~A. Lichtenstein,
\newblock ``Relevance of lung ultrasound in the diagnosis of acute respiratory
  failure: the blue protocol,''
\newblock {\em Chest}, vol. 134, no. 1, pp. 117--125, 2008.

\bibitem{volpicelli2012international}
G.~Volpicelli et~al.,
\newblock ``International evidence-based recommendations for point-of-care lung
  ultrasound,''
\newblock {\em Intensive care medicine}, vol. 38, no. 4, pp. 577--591, 2012.

\bibitem{dutta2019vgg}
A.~Dutta et~al.,
\newblock ``The {VIA} annotation software for images, audio and video,''
\newblock in {\em ACM Multimedia}, 2019.

\bibitem{simonyan2014very}
K.~Simonyan et~al.,
\newblock ``Very deep conv networks for large-scale image recognition,''
\newblock {\em arXiv:1409.1556}, 2014.

\bibitem{he2016deep}
K.~He et~al.,
\newblock ``Deep residual learning for image recognition,''
\newblock in {\em IEEE CVPR}, 2016, pp. 770--778.

\bibitem{lea2016temporal}
C.~Lea et~al.,
\newblock ``Temporal convolutional networks: A unified approach to action
  segmentation,''
\newblock in {\em ECCV}, 2016, pp. 47--54.

\end{thebibliography}

\end{document}